\documentclass[twocolumn,showpacs,preprintnumbers,amsmath,amssymb]{revtex4}

\usepackage{graphicx}
\usepackage{dcolumn}
\usepackage{bm}

\begin{document}


\title{Magnetic tuning of tunnel conductivity}

\author{Yu.G. Pogorelov}
\altaffiliation{Electronic address: ypogorel@fc.up.pt}
\affiliation{CFP and Departamento de F\'{\i}sica, Faculdade de
Ci\^{e}ncias, Universidade do Porto, Rua do Campo Alegre, 687,
4169-007 Porto, Portugal}%
\author{J.B. Sousa}
\altaffiliation{Electronic address: jbsousa@fc.up.pt}
\author{J.P. Ara\'{u}jo}
\altaffiliation{Electronic address: jearaujo@fc.up.pt}
\affiliation{IFIMUP and Departamento de F\'{\i}sica, Faculdade de
Ci\^{e}ncias, Universidade do Porto, Rua do Campo Alegre, 687,
4169-007 Porto, Portugal}%

\date{\today}

\begin{abstract}
Using the simplest two-subband Stoner model, it is shown that the
variation of the Fermi energy under applied magnetic field is inverse
proportional to the spontaneous magnetization and hence most pronounced
close to the critical Stoner condition, that is to the quantum critical
point of ferromagnetic transition. The perspectives of this result
for magnetic tuning of tunnel conductivity in spintronics devices
is discussed.
\end{abstract}



\maketitle

\noindent Development of new effective methods to control the spin 
degrees of freedom in electronic transport, generally referred to as 
spintronics, is one of the hottest topics in modern solid state physics. 
The earlier solutions in this area were intended on formation of a 
certain difference in kinetic coefficients for spin-up and spin-down 
electrons, controlled by the applied magnetic field (see e.g. recent 
review in \cite{aw}, \cite{das}), and they still encounter difficulties in 
achieving reasonable field effect, above several percent. An alternative, 
and more efficient, route exploits spin-dependent tunneling with controllable 
transparency of tunnel barrier \cite{mia}, \cite{moodera}, usually through 
the effect of mutual polarization of magnetic electrodes on the overlap 
matrix element \cite{jul}. With use of typical ferromagnetic metals, this effect 
amounts to some tens percent.

An intriguing possibility was indicated recently, consisting in tuning
the tunneling conductance through the magnetic field effect on the
barrier height \cite{shimada}, \cite{ono}, which in principle can be 
exponentially strong. One of possible realizations of this idea is related 
to the tunneling between two macroscopic, non-magnetic electrodes through 
an intermediate nanoscopic magnetic particle (island) embedded into the 
insulating spacer. The purpose is to control the tunnel conductance in such 
a N-I-M-I-N junction (where N are non-magnetic electrodes, I insulator
spacers, and M a magnetic metal nanogranule) by shifting the Fermi
level in the island by the applied magnetic field $H$.

\begin{figure}
\centering\includegraphics[%
  scale=0.7]{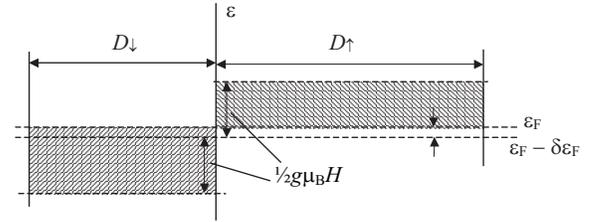}

\caption{\label{cap:fig0}Redistribution of Fermi electrons to a shifted Fermi 
level $\varepsilon_{\mathrm F} - \delta\varepsilon_{\mathrm F}$, after Zeeman 
shifts of each spin subband.}
\end{figure}

It was proposed by Ono \emph{et al} \cite{ono} that the field dependent
shift of Fermi level is 
\begin{equation}
\delta\varepsilon_{{\textrm{F}}}=\frac{1}{2}Pg\mu_{{\textrm{B}}}H,
\label{eq:00}
\end{equation}
where $P$ is the polarization of Fermi electrons 
$P=\left(D_{\uparrow}-D_{\downarrow}\right)/\left(D_{\uparrow}+
D_{\downarrow}\right)$ and $D_{\sigma}$ the respective partial Fermi 
densities of states. They considered the field effect due to Zeeman 
shifts of spin polarized subbands as shown schematically in Fig. 
\ref{cap:fig0}, where equal hatched areas correspond to redistribution 
of Fermi electrons to the shifted Fermi level. However, this simple 
argument does not take into account the field effect on the polarization 
of \emph{all the electrons}, which is essential for the overall number 
conservation and hence for the absolute position of Fermi level. Therefore, 
we feel a need in a more consistent treatment of the field effect on the 
overall band structure.

This can be done, for instance, using the simplest two-subband Stoner
model \cite{stoner} for the total electronic energy:
\begin{equation}
E = \sum_{\mathbf{k},\sigma}\varepsilon_{\mathbf{k}}n_{\mathbf{k},\sigma}
+J n_{\uparrow} n_{\downarrow}- h\sum_{\sigma}\sigma n_{\sigma},
\label{eq:0}
\end{equation}
\begin{figure}
\centering\includegraphics[%
  scale=0.7]{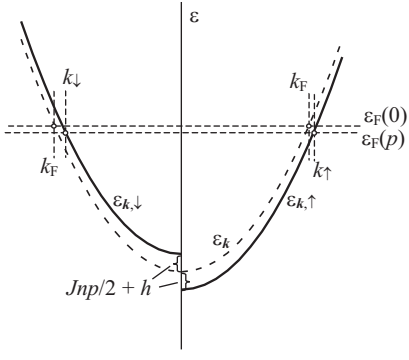}

\caption{\label{cap:fig1}Band structure of a Stoner ferromagnet displaying 
shifted subbands $\varepsilon_{k,\sigma}$ and shifted Fermi level 
$\varepsilon_{{\textrm{F}}}\left(p\right)$ at finite polarization of subbands, 
compared to the paramagnetic values $\varepsilon_{k}$ and 
$\varepsilon_{{\textrm{F}}}\left(0\right)$. Vertical dashed lines show displaced 
limiting wavenumbers $k_{\uparrow,\downarrow}$ for spin subbands, compared 
to the paramagnetic value $k_{\textrm{F}}$.}
\end{figure}

\noindent where $n_{\mathbf{k},\sigma}$ is the occupation number for electronic
state with momentum $\mathbf{k}$ and spin $\frac 1 2 \sigma$ (in what follows  
$\uparrow$ is identified with $\sigma = 1$ and $\downarrow$ with $\sigma = -1$), 
$n_{\sigma}= \sum_{\mathbf{k}}n_{\mathbf{k},\sigma}$ the partial electronic density, 
$\varepsilon_{\mathbf{k}}$ the paramagnetic dispersion law, $J$ the Stoner 
coupling constant, and $h=\frac {1}{2} g\mu_{{\textrm{B}}}H$ the Zeeman energy. 
The standard routine of this model is to minimize $E$ with respect to the 
\emph{band polarization} $p=\left(n_{\uparrow}-n_{\downarrow}\right)/n$ 
(for given total electronic density $n=n_{\uparrow}+ n_{\downarrow}$) and 
to obtain the equilibrium value of $p$, which can be field dependent: 
$p=p\left(h\right)$. But then we further use the obtained $p$ to calculate 
the shift of Fermi level $\varepsilon_{{\textrm{F}}}\left(p\right)$ vs its 
value in the paramagnetic state $\varepsilon_{\mathrm{F}}=
\varepsilon_{\mathrm{F}}\left(0\right)$ (see Fig. \ref{cap:fig1}).

For isotropic dispersion $\varepsilon_{\mathbf{k}}=\varepsilon_{k}$
and zero temperature, the occupation numbers are simply 
$n_{\mathbf{k},\sigma}=n_{k,\sigma}=\theta\left(k_{\sigma}-k\right)$,
where the limiting wavenumber $k_{\sigma}$ can be expressed from
the relation for the partial density:
\begin{eqnarray}
n_{\sigma} & = & \sum_{k<k_{\sigma}}1\approx\frac{1}{2\pi^{2}}
\int_{0}^{k_{\sigma}}k^{2}dk\nonumber \\
 & = & \frac{k_{\sigma}^{3}}{6\pi^{2}}=\frac{k_{\mathrm{F}}^{3}\left(1+
\sigma p\right)}{6\pi^{2}},
\label{eq:1}
\end{eqnarray}
through the paramagnetic $k_{\mathrm{F}}$ and $p$. Then the dispersion
law for (polarized) $\sigma$th subband can be defined in accordance with 
the usual scheme of Landau Fermi liquid (LFL) theory:
\begin{equation}
\varepsilon_{k,\sigma}=\frac{\delta E}{\delta n_{k,\sigma}}=
\varepsilon_{k}-\sigma\left(\frac{Jnp}{2}+h\right)+\rm{const},
\label{eq:2}
\end{equation}
where the $\sigma$-independent constant can be safely put zero. The uniform 
$\sigma$-dependent shift with respect to the paramagnetic dispersion 
$\varepsilon_{k}$ includes the two components of molecular field, 
the Stoner exchange field $Jnp/2$ and the external field $h$. The 
condition for unique Fermi level in both subbands reads
\begin{equation}
\varepsilon_{k_{\uparrow},\uparrow}=\varepsilon_{k_{\downarrow},\downarrow}
=\varepsilon_{\mathrm{F}}\left(p\right),
\label{eq:3}
\end{equation}
and, evidently, the paramagnetic Fermi level is 
$\varepsilon_{\mathrm{F}}=\varepsilon_{k_{\mathrm{F}}}$.
Then we present the total energy as
\[E=\sum_{\sigma}\sum_{k<k_{\sigma}}\varepsilon_{k}-np\left(\frac{Jp}{4}+
h\right)+{\textrm{const}},\]
and, taking account of Eq. \ref{eq:1} and using the parabolic dispersion
$\varepsilon_{k}\propto k^{2}$, obtain its explicit dependence on $p$:

\begin{eqnarray}
E\left(p\right) & = & n\left\{ \frac{3\varepsilon_{{\textrm{F}}}}{10}
\left[\left(1+p\right)^{5/3}+\left(1-p\right)^{5/3}\right]\right.\nonumber \\
 &  & \qquad\qquad\qquad\qquad\left.-p\left(\frac{Jp}{4}+h\right)\right\} .
\label{eq:5}
\end{eqnarray}
The necessary minimum condition for Eq. \ref{eq:5}:
\[\varepsilon_{{\textrm{F}}}\left[\left(1+p\right)^{2/3}-\left(1-
p\right)^{2/3}\right]=2\left(\frac{Jnp}{2}+h\right),\]
is just equivalent to Eq. \ref{eq:3}, and, being expanded for small
polarization $p\ll1$ and weak external field $h\ll Jnp$, leads to
\begin{equation}
p\left(h\right)\approx p\left(0\right)+\frac{27h}{8\varepsilon_{{\textrm{F}}}
p^{2}\left(0\right)},
\label{eq:6}
\end{equation}
where the spontaneous polarization 
\[p\left(0\right)=3\sqrt{\frac 1 2 \left(\frac {3Jn}{4\varepsilon_{{\textrm{F}}}}-1\right)}=
3\sqrt{\frac 1 2 \left(JD_{\mathrm F}-1\right)}\] 
includes the paramagnetic Fermi density of states 
$D_{\mathrm F}\approx \left(D_{\uparrow}+D_{\downarrow}\right)/2$. 
Finally, substituting Eq. \ref{eq:6} into Eqs. \ref{eq:3} and \ref{eq:2} results in 
the sought expression for shifted Fermi level:
\begin{equation}
\varepsilon_{{\textrm{F}}}\left(p\right)\approx\varepsilon_{{\textrm{F}}}
\left(0\right)\left[1-\frac{p^{2}\left(0\right)}{9}\right]-\frac{3h}{4p\left(0\right)}.
\label{eq:7}
\end{equation}
This formula indicates that the sharpest response of $\varepsilon_{{\textrm{F}}}$
to external field (when small enough: $h\ll\varepsilon_{{\textrm{F}}}p^{3}\left(0\right)$)
is achieved with the \emph{lowest} band polarization $p\left(0\right)$, that is close 
to the Stoner critical condition: $JD_{\mathrm F}\approx 1$. In this case, the 
$h$-dependent contribution to $Jnp/2$ from the second term in r.h.s. of Eq. \ref{eq:6} 
is $\approx h(3/2p(0))^2\gg h$, that is dominating over the proper Zeeman term $h$ 
in the molecular field $Jnp/2+h$, which defines a great difference with the 
Ono \emph{et al} result. This can be expected, considering the LFL relation 
$\varepsilon_{{\textrm{F}}}=\delta E/\delta n$ and the phenomenological 
dependence of total energy on external field $E = E_{H=0} -\chi H^2$ with the mean-field 
behavior of susceptibility $\chi \sim 1/p(0)$. We also notice that the band polarization 
$p$ in Eq. \ref{eq:7} is quite different from the Fermi level polarization $P$ in 
Eq. \ref{eq:00}, and (for non-parabolic dispersion) can even have opposite sign. 
Practically, the required closeness to critical condition can be met in various 
transition metal alloys, as e.g. Pd-Fe, Rh-Fe, or Ru-Co, though, when passing 
from bulk material to nanoparticles, the particular compositions of these systems 
might need to be reconsidered. 

Since the Stoner critical point corresponds to a kind of quantum phase 
transitions \cite{sachdev}, \cite{kirk}, an important question is the possible 
effect of quantum critical fluctuations which can put some restrictions 
on the choice of operating parameters. In particular, the correlation length of 
quantum fluctuations is estimated as $\xi_c \sim a k_{\mathrm B}T_c/
(\varepsilon_{\mathrm F}p(0))$ where $a$ is interatomic distance, $k_{\mathrm B}$ 
the Boltzmann constant, $T_c$ the characteristic temperature of magnetic ordering 
in the saturated state (far from the quantum critical point), and in order that 
fluctuations stay smaller of the nanoparticle size $d$ one needs the 
spontaneous polarization no less then $p_{min} \sim a k_{\mathrm B}T_c/
(d\varepsilon_{\mathrm F})$. Then, with typical values $a \sim 0.3$ nm, 
$T_c \sim 300$ K, $d \sim 5 nm$, $\varepsilon_{\mathrm F} \sim 1$ eV we obtain 
$p_{min} \sim 10^{-3}$, permitting a really high enhancement by Eq. \ref{eq:7}.

An additional enhancement of the field response of the N-I-M-I-N device can be found, 
considering Coulomb blocade and Coulomb staircase effects at tunneling
through nanoparticles \cite{shimada},\cite{ono},\cite{imamura}, so that the system 
can be previously driven by a common gate bias to the point of steepest $dI/dV$ and 
then used to detect the current variation due to weak magnetic signal
$\delta h$: 
\[\delta I=\frac{3\delta h}{4p\left(0\right)}\frac{dI}{dV}.\]

In conclusion, the simple analysis based on two-subband Stoner model for total 
electronic energy shows that an effective control of tunnel conductance can be 
achieved by tuning the position of Fermi level in one of the electrodes with 
relatively weak external magnetic field. This tuning is most pronounced when 
the material of the considered electrode is chosen to be close to the Stoner 
critical condition for ferromagnetic transition. The extension of this model 
to more realistic band structures and exchange coupling schemes \cite{marcus} 
is straightforward. The proposed magnetic tuning mode can open a new route for 
modern spintronic devices.

\end{document}